\documentclass[useAMS,usenatbib]{mn2e}
\usepackage{graphicx}
\usepackage[latin1]{inputenc}
\usepackage{color}
\usepackage{times}
\usepackage{natbib}
\usepackage{setspace}
\usepackage{amsmath}
\usepackage{subfigure}
\newif\ifAMStwofonts
\AMStwofontstrue
\definecolor{red}{rgb}{1,0.,0.}

\title[Galaxy morphologies at low redshift] {The
distribution of galaxy morphological types and the morphology-mass relation
in different environments at low redshift}
\author[Calvi et al.]{Rosa Calvi$^{1}$\thanks{E-mail:
email: rosa.calvi@unipd.it}, Bianca M. Poggianti$^{2}$, Giovanni Fasano$^{2}$, Benedetta Vulcani$^{1,2}$\\
  $^1$Astronomical Department, Padova University, Italy \\
  $^2$INAF-Astronomical Observatory of Padova, Italy\\}

\begin{document}

\date{Accepted ... Received ...}

\pagerange{\pageref{firstpage}--\pageref{lastpage}} \pubyear{}

\maketitle
\label{firstpage}

\begin{abstract}
We use $\sim$ 2000 galaxies belonging to different environments to
show how the fractions of different galaxy morphological types vary with global
environment and as function of galaxy stellar mass at low redshift.
Considering mass limited galaxy samples with
$ log_{10}M_{\star}/M_{\sun} \geq 10.25$, we find
a smooth increase/decline in the fraction of Es-S0s/late type
galaxies going from single galaxies, to binaries, to groups.
Considering all environments,
the fractional variation is more conspicuous for S0s
and late-types than for ellipticals solely
due to a sharp enhancement/dearth of S0s/late-types in clusters
compared to other environments.
%that the fraction of ellipticals
%not show a significant difference in clusters, groups, binary systems
%and isolated galaxy sample. On the contrary there is a larger
%variation for S0s whose fraction increases up to $\sim$50\% in more
%massive systems. The late-type fractions change strongly with
%environment above all in clusters where they fall down indicating a
%clear inversion with S0s.
%The morphology-mass relation shows no
%galaxy mass dependence of the morphological mix
%in each environment for masses $<10^{10.8-11}$, while at higher
%masses the variation is strong.
%At galaxy masses log$_{10}M_{\star}/M_{\sun} < 11$ there is also
%little variation of the
The morphological distribution of galaxies in the mass range
$ 10.25 < log_{10}M_{\star}/M_{\sun} < 11$ is rather independent
both of galaxy stellar mass and global environment, except in
clusters.
The morphologies of galaxies more massive than $log_{10}M_{\star}/M_{\sun}
= 11$ are instead a function of both galaxy mass and global environment.
The morphology-mass relation therefore changes with global environment,
showing that galaxy stellar mass cannot be the only parameter
driving the morphological distribution of galaxies.
The  morphology-mass relations for S0 and late-type galaxies in clusters
are peculiar compared to other environments, and this strongly suggests
that cluster-specific effects act on these
two types of galaxies, and that a significant number of S0s in clusters
has  a different origin with respect to S0s in other environments.
\end{abstract}


\section{Introduction}
\label{s:introduction}
Morphology is one of the key observables in extragalactic astronomy
for understanding the formation and evolution of galaxies, and it is
the consequence of all physical processes at work.

The origin of such processes can be both external (environmental) or
internal (intrinsic) to the galaxy itself.  The strongest evidence of
a correlation between the morphological types (Hubble types) and the
environment in which galaxies are located comes from studies conducted
on galaxy clusters.  The well-known morphology-density relation, first
described by \citet{D1,D2} shows that ellipticals and S0s
are the dominant population in high density regions while galaxies in
low density regions are mainly spirals.  A relation between galaxy
morphologies, or structural parameters, and local density was found to
exist also in galaxy groups by \citet{P} and \citet{T}, and in the
SDSS general field \citep{G,K}.

Studies of higher redshift clusters proved the existence of a
morphological evolution of galaxies \citep{D3,F1,PP} with an increase of
the S0 fraction at low redshift at the expense of the spiral
population.
%In clusters, this evolution has taken place more recently
%in intermediate density regions than in high density regions \citep{S},
%and has mostly occurred at $z<0.5$ \citep{D}. In addition,
The morphological evolution depends strongly on galaxy
stellar mass both in clusters and in the field
\citep{V,O}, and it is
stronger in low-mass than in massive clusters \citep{PO,W,J}. Measuring the evolution of 
galaxy morphologies as a function of
environment represents an important step towards a better
understanding of the observed morphological mix, and for disentangling
between the role of environment and the role of galaxy intrinsic
effects. \citet{KK} found that galaxy structure is
strongly correlated with galaxy stellar mass. On the other hand, the
distribution of galaxy stellar masses itself depends on local density
\citep{B,BO,VV}.
Therefore it is hard to assess the relative roles played by
the galaxy mass and by the physical processes linked to the
environments that the galaxy experienced during its life-time.

To this aim, in the last decade, a great effort has been made to
obtain large catalogues of morphologically classified galaxies.  For
example, detailed morphological classifications have been obtained by
\citet{F2} for cluster galaxies at low redshift in the
WIde-field Nearby Galaxy-cluster Survey
(WINGS), and several catalogues with visual classifications have been
produced for SDSS field galaxies \citep{N,BA,L,FU}. 

Recently, \citet{BM}, using a local, visually classified
sample from the Galaxy Zoo project, analysed the relationship between
galaxy morphology, colour, environment and stellar mass characterising
the environment by local galaxy density and by distance from the
nearest group/cluster with $M = 10^{13}- 10^{15} \, h^{-1} \,
M_{\odot}$.
%They found that both the morphological and the color
%distributions vary with stellar mass and that, at fixed stellar mass,
%morphology displays weaker environmental trends than color.  However,
They showed that ``only a small part of both the morphology-density and
colour-density relations can be attributed to the variation in the
stellar-mass function with environment.''

%and that the fraction of
%red galaxies varies more strongly than the early-type fraction, both
%being higher in denser environments.
%
%Moreover, they observed that red
%spirals with low mass have a peak at intermediate local density
%meaning that some process occurs in the high density environments
%which influence this fraction while galaxies with higher mass, both
%spirals and early-type, are less dependent from the environment.
% and this trend is very similar to those found for local density

In spite of the recent observational progress, what drives the
observed trends of the different morphological types remains still a
controversial topic not fully explored.
For example, to date, there has not been any large statistical study
on the morphological distribution of galaxies with {\it global}, as opposed
to local, environment. By galaxy ``global environment'' we mean the
structure to which the galaxy belongs,
from massive clusters, to groups, to haloes hosting a single luminous galaxy,
over the whole range of
galaxy system masses, and corresponding dark matter halo masses.

This letter is the first work which
analyses how the morphological mix
and the relation between morphological fractions and galaxy stellar mass
change in the local Universe not as a function of local galaxy density, but
as a function of global environment.\footnote{See also \citet{W2}
submitted, Dave Wilman 2011 private communication.}
To do this we use two galaxy datasets representative of the whole range
of cosmic structures: the Padova-Millennium Galaxy and Group Catalogue (PM2GC)
of groups, binary systems
and isolated galaxies \citep{C}, built from the Millennium Galaxy
Catalogue (MGC) \citep{LI},
and the cluster sample from the WINGS survey \citep{F2}.

% we obtained isolated, binary
%systems and group galaxy samples, which together represents the
%general field (PM2GC catalogues, Calvi et al. 2011), and as comparison
%we used a sample of galaxies taken from WINGS survey (Fasano et al. 2006)
%an homogeneus survey of galaxy clusters (see \S2).

The higher quality imaging of the MGC and WINGS surveys compared to
the Sloan survey and the combination of multiwavelength photometry and
spectroscopy for both our samples allow us to give a clear
representation of the morphological distribution of galaxies in
different environments as a function of galaxy stellar mass.
%Our results may be useful for a comparison with studies of the
%morphological evolution of galaxies (Oesch et al. 2010, Cassata et
%al. 2010) and the morphology-density relation in the field at high redshift
%(Tasca et al. 2009, Capak et al. 2007). In $\S$2 we present
%the PM2GC dataset, give an overview of the WINGS survey and of the
%methods used to derive galaxy stellar masses and morphological types.
%In $\S$3 we show the results, and finally a brief summary is
%presented in $\S$4.
We adopt $H_{0}=70 \, \rm km \, Mpc^{-1} s^{-1}$,
$h=H_{0}/$100, $\Omega_{m}=$0.3 and $\Omega_{\lambda}=$0.7.
\section{The Datasets}
The data used in this letter consist of two samples of galaxies in
the local Universe, in which we identify different environments.

We selected group, binary, isolated and general field galaxies from
the PM2GC \citep{C}, a sample sourced from the MGC
\citep{LI,DR}. The MGC is a B-band contiguous survey of
$\sim$38$\rm deg^{2}$, complemented by a highly complete spectroscopic
survey down to B=20.  The PM2GC is based on the spectroscopically
complete sample including all galaxies in the MGC area in the
redshift range 0.03$\leq$z$\leq$0.11 brighter than M$_{B}<$-18.7.

The PM2GC catalogue of groups (PM2-G) contains 176 galaxy groups at
0.04$\leq$z$\leq$0.1 with at least three members brighter than
M$_{B}=$-18.7, identified using a FoF group-finding algorithm. A
galaxy is a group member if its spectroscopic redshift lies within
$\pm 3\sigma$ from the median group redshift and if it is located
within a projected distance
of $\leq 1.5 \, R_{200}$\footnote{$R_{200} = 1.73 \, \frac{\sigma}{1000 \, \rm km \, s^{-1}} \, {\frac{1}{\sqrt{{\Omega}_{\Lambda} + {\Omega}_{0}(1+z)^3}}} \, h^{-1} \, \rm Mpc$} from
the group geometrical centre. Galaxies which have no neighbour or
solely one with a projected mutual distance $\leq 0.5 h^{-1}$Mpc and a
redshift within 1500$\rm km \, s^{-1}$ are considered ``isolated''
galaxies (PM2-FS) or ``binary-system'' galaxies (PM2-FB),
respectively. The general field (PM2-GF) includes all galaxies regardless
of environment, and assembles galaxies in the PM2-G, PM2-FS and PM2-FB
together with galaxies that, although located in a trial group, did
not make it into the final group sample.  A detailed description of
the FoF method and catalogues of PM2GC groups and galaxies in
different environments can be found in \citet{C}.

For the analysis conducted in this letter we prune the PM2GC catalogues of
isolated, binary system and general field galaxies to the same
redshift range of groups, thus 0.04$\leq$z$\leq$0.1.

In addition, we use a sample of galaxies belonging to 21
clusters taken from WIde-Field Nearby Galaxy-cluster Survey (WINGS)
\footnote{http://web.oapd.inaf.it/wings}, a multiwavelength survey
based on deep optical (B,V) wide field images of 77 clusters at
0.04$<$z$<$0.07 selected in the X-ray \citep{F2} which span a
wide range in velocity dispersion ($\sigma$ typically between 500-1100
km s$^{-1}$) and X-ray luminosity (L$_{X}$ between
0.2-5$\times10^{44}$ergs$^{-1}$). The galaxy spectroscopic
completeness of this subsample of clusters
is $\geq$50\%, and a spectroscopic incompleteness correction
has been applied to our analysis. For redshifts measurements, cluster
membership and completeness see \citet{CV}.

Although reliable estimates of individual halo masses are available
only for WINGS structures based on cluster velocity dispersions
and X-ray luminosity, by combining PM2GC and WINGS certainly we are
statistically investigating a very broad range of global environments,
and corresponding dark matter halo masses, from $\sim 10^{15}
M_{\odot}$ for WINGS, to the order
of $\sim 10^{12} M_{\odot}$ or lower for single galaxies.

For all galaxies in PM2GC and WINGS, stellar masses have been estimated
using the \citet{BL} relation which correlates the
stellar mass-to-light ratio (M/L) with the optical colors of the integrated
stellar populations\footnote{The relation is:
$\log_{10}(M_{\star}/L_{B})=a_{B}+b_{B}(B-V)$ having considered the
B-band photometry, a Bruzual \& Charlot model with $a_{B}$=-0.51 and
$b_{B}$=1.45 for a Salpeter IMF (0.1-125 M$_{\odot}$), subsequently scaled to
a Kroupa (2001) IMF, and solar metallicity.}.
Our mass estimates are in agreement with DR7 masses and with estimates
obtained from the spectrophotometric model fitting of WINGS spectra,
with a scatter that is equal to the typical error on photometric mass
estimates (0.2-0.3dex), without any dependence on morphological type
(see Calvi et al. 2011 for PM2GC, and Vulcani et al. 2011 and
Fritz et al. 2011 for WINGS for all details regarding stellar masses).
\begin{table}
%\scriptsize{
\centering
\resizebox {0.51\textwidth }{!}{
\begin{tabular}{lccccc}
\hline
%\textbf{Environment} & \multicolumn{2}{|c|}
\textbf{Envir.} & \textbf{N. of gal.} & \multicolumn{4}{c}{\textbf{Galaxy type}}\\
%            &  $M_{\star}\geqslant 10^{10.25}M_{\odot}$& & & & \\
            & & Ellipticals & S0s & Late-type & Early-type \\
\hline
WINGS  & 690 (1056.12) & 33.8$\pm$1.5\% & 50.7$\pm$1.5\% & 15.4$\pm$1.0\% &
            84.5$\pm$1.0\%\\
PM2-GF & 1188 & 27.0$\pm$1.3\% & 28.7$\pm$1.3\% & 44.3$\pm$1.5\% &
            55.7$\pm$1.5\%\\
PM2-G & 583 & 29.7$\pm$1.9\% & 32.4$\pm$2.0\% & 37.9$\pm$2.1\% &
            62.1$\pm$2.5\%\\
PM2-FB & 174 & 25.3$\pm$3.5\% & 25.8$\pm$3.6\% & 48.8$\pm$4.0\% &
            51.1$\pm$4.0\%\\
PM2-FS & 334 & 21.5$\pm$2.3\% & 24.2$\pm$2.5\% & 54.2$\pm$2.8\% &
            45.7$\pm$3.0\%\\
\hline
\end{tabular}}
\centering
\caption{Number of galaxies and fractions of each morphological type
  in the PM2GC and WINGS mass-limited samples with
  $M_{\star}$=10$^{10.25}M_{\odot}$. Early-type galaxies comprise ellipticals and S0s.
  Errors are binomial. The WINGS number of galaxies between
  brackets is the weighted total number of galaxies above our completeness
  limit.}\label{t2}
\end{table}
In order to properly investigate the relation between morphology, galaxy mass
and environment,
%environmental dependence of the
%morphological fractions and the relation between morphologies and
%galaxy masses,
without being affected by the different distributions
of star formation histories (hence colors) in different environments,
it is important to restrict the analysis to a galaxy sample that is complete in
mass. The PM2GC galaxy stellar mass completeness limit
was computed as the mass of the reddest $M_{B}$=-18,7 galaxy ($B-V$=0.9) at
our redshift upper limit (z=0.1), and it is equal to
$M_{\star}$=10$^{10.25}M_{\odot}$.
The WINGS sample is complete in mass down to $M_{\star}$=10$^{9.8}M_{\odot}$,
but for the purposes of this work, we have adopted for WINGS the same
galaxy mass limit as for the PM2GC.

Hence, all the results presented in this paper are based on galaxy stellar
mass limited samples ($M_{\star} \geq 10^{10.25}M_{\odot}$),
for a \citet{KR} IMF.
The total numbers of galaxies in each environment
above our mass limit are listed in Table \ref{t2}.

To morphologically classify galaxies in both samples we used MORPHOT,
an automatic tool purposely designed \citep{F4}, see also
Appendix A in \citet{F3}, which used B-band images for
PM2GC galaxies and V-band images for WINGS. Visual inspection of
a random subset of 300 galaxies
showed no significant systematic shift in broad morphological
classification (E, S0 or late-type) and a typical scatter ($\Delta T$=1.2)
between the V and B WINGS images. By adding to the
classical CAS (Concentration/Asimmetry/clumpinesS) parameters a set of
additional indicators derived from digital imaging of
galaxies, this tool mimics the visual classifications.
The MORPHOT morphological classifications have been proved to
be as effective as the eyeball estimates,
as shown by the comparison with visual classifications of
WINGS and SDSS images
yielding an average difference in Hubble type $\Delta T$
($ \leq 0.4$) and scatter ($\leq 1.7$)
comparable to those among visual classifications
of different experienced classifiers
(see Fasano et al. 2011).
In particular, MORPHOT has been shown to be
able to distinguish between ellipticals and S0 galaxies with
unprecedented accuracy.
%, with a scatter
%that is comparable to the average scatter among visual classifications
%of different experienced classifiers \citep{F4}.
\section{Results}
\subsection{The Hubble type fractions}
\begin{figure}
	\vspace{-2cm}
 	\includegraphics[scale=0.36]{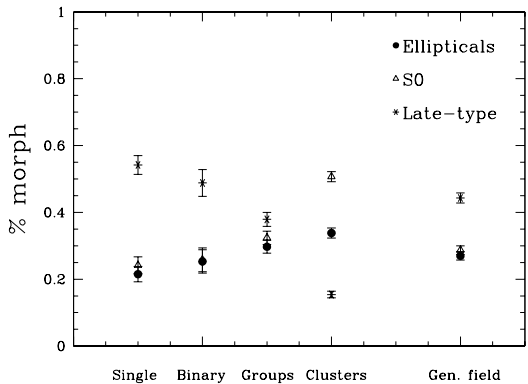}
 	\caption{Morphological fractions in different global
environments for our mass-limited samples (\textbf{for $log_{10}M_{\star}/M_{\sun}
\geq 10.25$)}, as in Table \ref{t2}. Errors are binomial.}
\label{F1}
\end{figure}
In Table \ref{t2} and Figure \ref{F1}
we show the morphological fractions for the PM2GC and
WINGS mass-limited samples for galaxies with $log_{10}M_{\star}/M_{\sun}
\geq 10.25$.
%{\it Se c'e' bisogno di spazio possiamo togliere questa frase, senno' possiamo
%lasciarla. We note that the general field (PM2-GF) is
%dominated by group galaxies, that represent about half of the
%general field population. The PM2-GF fractions are clearly an
%average of those in the different field environments.}

The fractions of ellipticals, S0s and early-type galaxies (E's + S0's)
progressively increase in more massive environments, from single
galaxies to binary systems, to groups and clusters.
Late-type galaxies follow the opposite trend with global environment.

We note that elliptical galaxies have the smallest fractional
variation with environment. Their fraction in clusters ($\sim$34\%)
is not very dissimilar from that in the general field ($\sim$27\%) .
For isolated galaxies, binary systems and groups,
%separately, we find that,
although the trend is slowly increasing as we
move towards more massive environments, the fraction of ellipticals
is always between $\sim$22-30\%.

%In contrast, there is a much larger variation of the S0 fraction
%with environment,
%ranging from 24\% among single galaxies,
%to 32\% in groups, with a jump to 50\% in clusters.
%S0's in clusters are almost twice more common than in the general field.
%In all environments but clusters, the frequency of S0s is very similar to
%Es ($\sim$1:1 ratio), while in clusters this ratio is 1.5.

The fraction of S0s is very similar to the elliptical fraction
in all environments ($\sim$1:1 number ratio) except in clusters (1.5).
The S0 fraction ranges from 24\% among single galaxies,
to 32\% in groups, with a jump to 50\% in clusters.
S0's in clusters are almost twice more common than in the general field.

%The trends are outstanding also when considering all early-type
%galaxies together, whose fraction ranges from 46\% among single
%galaxies, to 62\% in groups, to 85\% in clusters, and 56\% in the
%general field.
As a consequence, the early-type fraction ranges from 46\% among single
galaxies to 85\% in clusters.
It is striking that even in the least massive
environments we consider, (those of single galaxies),
%the population of
%galaxies more massive than our mass limit is composed of a large
%number of early-type galaxies representing
almost half of the total galaxy population above our mass limit is composed
of early-type galaxies.

Correspondingly, the fraction of late-type galaxies changes strongly with
environment.  For PM2-FS and PM2-FB this fraction is about $\sim$50\%,
it decreases for group galaxies to $\sim$38\%, and then it declines sharply
in clusters (mirroring the S0 cluster enhancement)
reaching a fraction of $\sim$15\%, a third with respect to the
general field ($\sim$44\%).

In \citet{PO}\footnote{In the 2009 work the analysis was
carried out for a magnitude-limited sample of galaxies. We have
verified that the results do not change for the mass-limited WINGS
sample we use in this paper.} we found that in clusters
($\sigma > 500 \rm \, km \, s^{-1}$)
there is no
correlation between the fractions of late-types, S0s and ellipticals
and cluster velocity dispersion or X-ray luminosity (both of
which are cluster mass proxies), except for a weak anticorrelation of
the late-type fraction with X-ray luminosity.
%This holds over the
%WINGS range $\sim 500 < \sigma < 1100 \rm \, km \, s^{-1}$ and $0.2 <
%L_X < 5 \times 10^{44} \rm \, erg \, s^{-1}$.
Together with the results of this paper, this suggests
%a ``saturation'' of the morphological mix in global environments at the
%cluster level.  T
that the morphological trends we observe going from less
massive (single galaxies) to more and more massive environments
(groups and clusters) must flatten out for the most massive haloes.

In summary, we see a smooth increase/decline in the fraction of
Es-S0s/late type galaxies going to more and more massive
environments. On top of that, the cluster environment is characterized
by a sharp enhancement/dearth of the S0/late-type fractions compared
to other environments. For clusters above $500 \rm \, km \, s^{-1}$
the morphological fractions do not depend on halo mass anymore.
Overall, the fraction of elliptical galaxies
%does not depend very strongly on global environment,
represents always at least 20\% of
the galaxy population even in the least massive environments, and at
most 34\% in clusters.

\subsection{The Morphology-Mass relation in different environments}
In this section we investigate the dependence of the distribution of
morphological types on galaxy stellar mass, and whether this changes
with global environment. Table \ref{t3} lists the fractions of each morphological type at masses $10.25\leq$ 
log$_{10}M_{\star}/M_{\sun}<11.0$ and log$_{10}M_{\star}/M_{\sun}\geq 11.0$
in different environments and can be used in the following to compare
the morphological fractions at low- and high-masses, and in different 
environments in the same mass range.
\begin{table}
%\scriptsize{
\centering
\resizebox {0.51\textwidth }{!}{
\begin{tabular}{lccccc}
\hline
%\textbf{Environment} & \multicolumn{2}{|c|}
\textbf{\% morph.type} & \textbf{WINGS} & \textbf{PM2GC-GF} & \textbf{PM2GC-G} & \textbf{PM2GC-FB} & \textbf{PM2GC-FS}\\
\hline
 ell $M\geq 10^{11}M_{\odot}$ & 63.2$\pm$4.4\% & 42.0$\pm$4.5\% & 48.2$\pm$4.5\% & 29.4$\pm$13.5\% &
            26.6$\pm$9.6\% \\
 ell $M< 10^{11}M_{\odot}$ & 29.3$\pm$1.6\% & 24.8$\pm$1.5\% & 26.6$\pm$2.0\% & 24.8$\pm$3.7\% &
			21.0$\pm$2.5\% \\
 s0 $M\geq 10^{11}M_{\odot}$ & 26.3$\pm$4.0\% & 20.0$\pm$3.5\% & 21.6$\pm$5.0\% & 29.4$\pm$13.5\% &
            13.3$\pm$7.8\% \\
 s0 $M< 10^{11}M_{\odot}$ & 54.4$\pm$1.7\% & 30.0$\pm$1.5\% & 34.2$\pm$2.0\% & 25.4$\pm$3.8\% &
            25.3$\pm$2.6\% \\
 lt $M\geq 10^{11}M_{\odot}$ & 10.4$\pm$2.9\% & 38.0$\pm$4.5\% & 30.1$\pm$6.0\% & 41.1$\pm$14.3\% &
            60.0$\pm$10.4\% \\
 lt $M< 10^{11}M_{\odot}$ & 16.1$\pm$1.2\% & 45.2$\pm$1.5\% & 39.2$\pm$2.5\% & 49.6$\pm$4.3\% &
            53.6$\pm$3.0\%\\
\hline
\end{tabular}}
\centering
\caption{Fractions of each morphological type above and below $M= 10^{11}M_{\odot}$ in
different environments. Errors are binomial.}\label{t3}
\end{table}
\subsubsection{Fixed environment}

%The differences in morphological fractions at masses below and
%above log$_{10}M_{\star}/M_{\sun} = 11.0$  are summarized in Table~2
%and are highly significant in clusters, groups and general field.
The morphological fractions at masses below and above
log$_{10}M_{\star}/M_{\sun} = 11.0$ are different (Table \ref{t3}). These
differences are statistically significant in clusters, groups and
general field, while among binaries and single galaxies numbers are
too low to draw conclusions.

\begin{figure}
	\vspace{-3pt}
 	\includegraphics[scale=0.4]{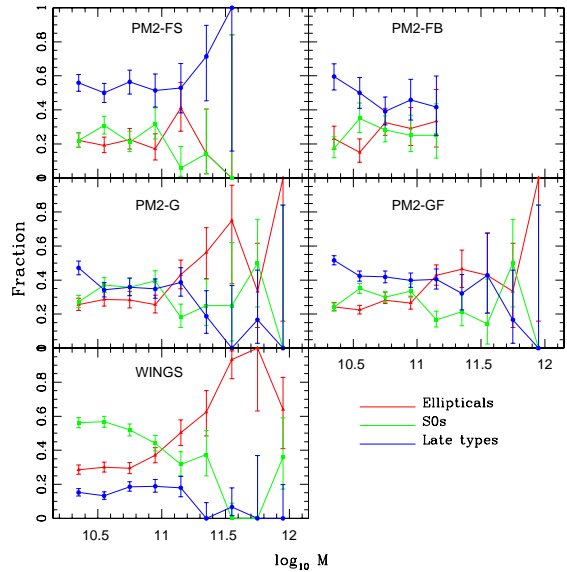}
 	\caption{The fraction of each
 	morphological type as a function of stellar mass in different
 	environments: PM2 single-field (top left), PM2 binary-field
 	(top right), PM2 galaxy groups (middle left), PM2 general
 	field (middle right), WINGS (bottom left, reproduced from Fig.10 of Vulcani et al. 2011).
    The WINGS number of galaxies is weighted on the total number
    of galaxies above our completeness limit. Errors are binomial.}
\label{F8}
\end{figure}

Figure \ref{F8} shows how the fraction of each morphological type
varies with stellar mass in the four PM2GC environments and in
clusters. We note that, as discussed in detail in Calvi et al. (in prep.),
in binary and single systems there are no galaxies with
masses log$_{10}M_{\star}/M_{\sun} \geqslant 11.2$ and
log$_{10}M_{\star}/M_{\sun} \geqslant 11.55$, respectively, indicating
that the most massive galaxies are only present in the most massive
haloes. Up to log$_{10}M_{\star}/M_{\sun} \sim 11.1$, in the PM2GC there is generally
little dependence of the morphological fractions on galaxy mass.
Moreover, at these masses
late-type galaxies tend to be the most common type of galaxies
in all PM2GC environments, being matched by S0s only in groups.

In contrast, at masses log$_{10}M_{\star}/M_{\sun} > 11.1$, there is a
noticeable dependence of the morphological mix on galaxy mass in those
environments hosting galaxies this massive.  In the PM2-FS the
late-type galaxy fraction rises towards higher masses, while in groups
it declines, as the most massive group galaxies are mostly
ellipticals. 
%In the PM2-GF, where the fractions are mediated over all
%environments, the mass dependence is quite weak at all masses.

%, with a
%tendency for the elliptical fraction to increase and the S0 fraction
%to decrease for masses in the range log$_{10}M_{\star}/M_{\sun} =
%11-11.6$.

In clusters, from \citet{V},
the trends partly resemble those in the
PM2GC groups, except that in clusters the dominance of ellipticals at
high masses is even more pronounced and for log$_{10}M_{\star}/M_{\sun} >
10.8$ there is a clear declining
trend of S0s with galaxy mass that is absent in groups.
This behaviour of the S0 fraction-galaxy mass relation in clusters is
different from all other environments,
%. S0 galaxies in clusters become
%less and less frequent as we go to higher galaxy masses, while in
%other environments
where the S0 fraction remains rather constant at all galaxy masses.

%the fraction of
%ellipticals is flat up to log$_{10}M_{\star}/M_{\sun} \sim 11$, then
%it steeply increases at higher masses dominating over the other types.
%S0s dominate in the range log$_{10}M_{\star}/M_{\sun} = 10.3- 11$, and
%their fraction decreases with mass for log$_{10}M_{\star}/M_{\sun} >
%10.8$.  The fraction of late-type galaxies is flat up to
%log$_{10}M_{\star}/M_{\sun} \sim 11.2$, then it is consistent with
%zero at higher masses.  These

%, with S0s being by far the
%most common type of galaxies for log$_{10}M_{\star}/M_{\sun} < 11$ in
%clusters.

Though it is commonly believed that the morphological mix depends
on galaxy mass, our results highlight that such a dependence
is generally weak at masses in the range
log$_{10}M_{\star}/M_{\sun} = 10.25-11$, while it becomes strong
at higher masses, where the dominant galaxy type changes with
global environment (elliptical in clusters and groups, late-type among
single galaxies).

%Moreover, the morphological trends with mass at high masses
%change with global environment,
%This change is
%noticeable going from clusters/groups where the high mass end is
%dominated by ellipticals, to single galaxy environments where the
%majority of massive galaxies are late-type.
%Moreover, the behaviour of the S0 fraction-galaxy mass relation in clusters is
%different from all other environments. S0 galaxies in clusters become
%less and less frequent as we go to higher galaxy masses, while in
%other environments their fraction remains rather constant at all galaxy masses.
\subsubsection{Fixed morphological type}
\begin{figure}
	\vspace{-10pt}
 	\includegraphics[scale=0.48]{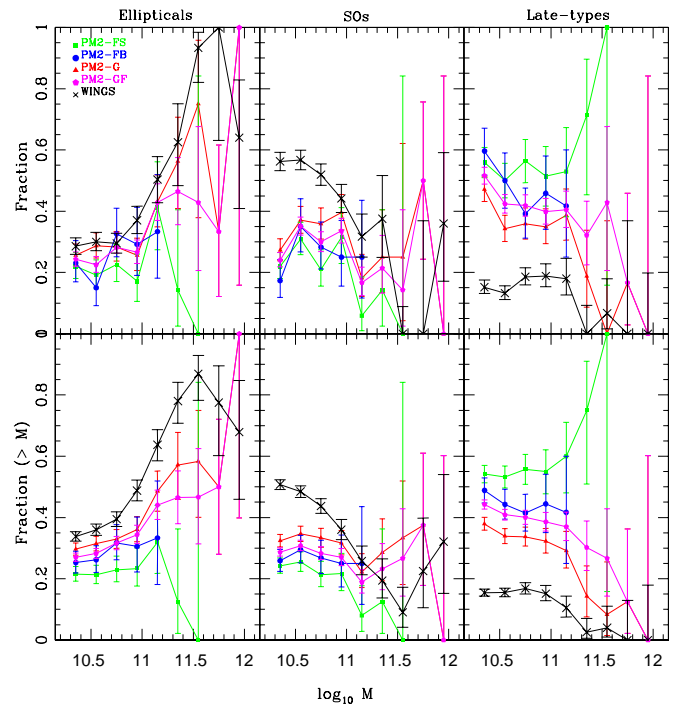}%{MMR_new.eps}
 	\caption{{\bf Top.} The fractions of elliptical (left panel), SO
 	(central panel) and late-type (right panel)
 	galaxies in different environments as a function of stellar mass. {\bf Bottom.} Cumulative distributions of ellipticals, S0s and late-type galaxies: fractions of each type among galaxies more massive than log$_{10}$M on the X axis. This plot can
used to infer the global morphological fractions in mass-limited samples
for any mass limit higher than ours. Errors are binomial. (The WINGS galaxies are weighted on the total number
    of galaxies above our completeness limit.)}
\label{F10}
\end{figure}
Figure \ref{F10} comprises all environments, and directly shows
the strong environmental variation of the morphology-mass relation
for each galaxy type.
The upper panels present the differential fractions,
and the lower panels the cumulative distributions showing the fractions
for galaxies more massive than a mass $M$.
%Lower panels can thus be
%used to know the global morphological fractions in mass-limited samples
%for any mass limit higher than ours.
Comparing our general field with results obtained using the
\citet{N} catalogue at the redshifts of our work, we
find qualitatively similar morphological trends with mass, though
even steeper high-mass slopes for ellipticals and late-types.

The top panels in the figure highlight that the variation of the
morphology-mass relation {\it with environment} is pronounced
(1) above all at high galaxy masses (log$_{10}M_{\star}/M_{\sun} >11$),
but also
(2) at low galaxy masses for S0 and late-type galaxies in clusters
compared to the other environments.

(1) In non-cluster environments, the fraction of each morphological type
at masses $ 10.25 < log_{10}M_{\star}/M_{\sun} < 11$
changes relatively little from one environment to the other,
while it varies much more strongly with environment for
$log_{10}M_{\star}/M_{\sun} >11$ (also Table \ref{t3}). The environmental variation at
high masses is especially outstanding for ellipticals and for late-types.

(2) The peculiarity of the behaviour of the S0 and late-type fractions in
clusters suggests cluster-specific effects on these two types of
galaxies.  It also suggests that a significant number of S0 galaxies
in clusters has a different origin with respect to S0s in other
environments.  The S0 fraction-galaxy mass relation in clusters
resembles more closely the late-type fraction-mass relation in the
general field and groups than the S0-mass relation in non-cluster
environments, in agreement with a scenario in which clusters are very
efficient in transforming spiral galaxies into (some of today's) S0s
\citep{D3,F1}. We speculate that
while a large number of cluster S0s have a ``late-type origin'',
a large number of non-cluster S0s have a truly ``early-type origin''
(a dominant spheroidal component that forms a very small disk, an
``intermediate type'' between disky ellipticals and Sa's).
%,
%given also that the S0 fraction outside of clusters follows so closely
%the elliptical fraction.
% and the trends with galaxy mass are not so dissimilar.

%The fact that the S0-mass
%relation is much flatter in the general field, groups and binary
%systems supports the hypothesis that the S0 population is formed
%through at least two channels depending on their mass and environment.

%The late-type galaxies are the dominant population in these kind of
%environments above all at low masses and in least massive environments. The
%fractions of ellipticals and S0s are smaller at low mass but increase
%at high mass. However,

%At each given galaxy mass, %we can certainly state that
%there is a general decrease in the elliptical and S0 fraction going to less
%and less massive environments, and a corresponding increase of
%the late-type fraction, with environmental differences being more
%conspicuous at higher galaxy masses.

The fact that the morphology-mass relation changes significantly with
environment
demonstrates that galaxy mass is not the only parameter driving the
morphological distribution of galaxies.
%The latter depends on galaxy mass only at high galaxy masses, but in
%a way that varies with global environment.
Thus, the change of morphological mix with global
environment cannot be solely due to an eventual change in the galaxy mass
function with halo mass (see Calvi et al. in prep. for a description
of the galaxy mass function as a function of global environment).
%a detailed
%analysis of the latter).
\section{Summary}
We use group, binary and isolated field galaxies from the PM2GC sample
and cluster galaxies from the WINGS survey to study the morphological
fractions and morphology-mass relation in different environments for
mass-limited galaxy samples with $ log_{10}M_{\star}/M_{\sun} \geq
10.25$.  Morphologies were derived from B-band and V-band images in
PM2GC and WINGS, respectively, with our tests showing no systematic
shift in broad morphological classification between B and V.

%In this letter we have investigated
%%addressed two of the main questions concerning the
%%morphologies of galaxies:
%how the distribution of
%morphological types and the morphology-mass relation change in
%different global environments. To do this we have considered
%mass limited samples of galaxies at low redshift
%in different %global environments representing a wide range of
%cosmic structures:
%galaxy groups, binary systems and
%isolated galaxies, assembled in a general field, and clusters.
%We summarize our results as follows:
\begin{itemize}
\item
We observe a smooth increase/decline in the fraction of Es-S0s/late type
galaxies going from single galaxies, to binaries, to groups.
% and clusters.
Considering all environments, the fractional variation
is more conspicuous for S0s
and late-types than for ellipticals,
due to a sharp enhancement/dearth of S0s/late-types in clusters
compared to other environments. \citet{PO} found that
above $500 \rm \, km \, s^{-1}$
the morphological fractions do not depend on halo mass anymore.
%The fraction of ellipticals does not depend very strongly on
%global environment.
%The total fraction of ellipticals is quite similar in all
%environments.
%The fraction of S0s ranges from 24 to 32 \%  in single,
%binary, general field and group environments,
%but is much higher in clusters ($\sim$ 50\%) where this morphological
%type is the most common.
%The fraction of late-type galaxies is very low
%in clusters ($\sim$16\%) compared to groups, binary systems and isolated
%sample ($\sim$38\%, $\sim$49\%, $\sim$54\%) and we observe a very clear inversion with S0s.
\item %We show the morphology-mass relation in different environments.
In all environments
the morphological fractions strongly depend on galaxy stellar mass {\it only}
for masses above log$_{10}M_{\star}/M_{\sun} \sim 11$ (10.8 in clusters),
while they are weakly dependent
of mass at  $ 10.25 < log_{10}M_{\star}/M_{\sun} < 11$.
At high galaxy masses, the dominant type
changes with global environment (elliptical in clusters and groups,
late-type in single galaxies).
\item Also the variation of the morphology-mass relation with environment
is much more pronounced at log$_{10}M_{\star}/M_{\sun} > 11$, especially
for ellipticals and late-types, while at lower masses there is relatively
little change from one environment to the other, except for clusters.
\item The morphology-mass relations for cluster S0s and late-types
remarkably differ from the corresponding relations in all other
environments.
%S0 galaxies in clusters become less and less frequent at higher galaxy
%masses, while in other environments their fraction remains rather constant
%at all galaxy masses.
Our findings strongly suggest that cluster-specific effects act on these
two types of galaxies, and that a significant number of S0s in clusters
has  a different origin with respect to S0s in other environments.

Our results show that the morphology-mass relation changes with global
environment and that galaxy stellar mass cannot be the only
parameter driving the morphological distribution of galaxies.

%we observe that the fraction
%of each type does not show significant changes at low masses in groups,
%binary systems and isolated sample while for clusters we observe a
%very steep increase in the fraction of S0s going to lower masses.
%The SOs galaxies in clusters for the most part have low masses
%and it is noted that the trends of S0s and late-types show an inversion
%passing from less dense environments to denser environments.

%In clusters the fraction of S0s at any given mass
%is always higher than in groups, binary
%systems and the isolated sample, and thus also the general field.
%except at very high masses ($>11.5$). {\textit The late-type galaxies tend to have not very high
%mass in clusters and their fraction at any given mass is always lower
%than in other environments. In other
%environments the dominant type is the late-type which can reach very
%high masses only in the single sample.}

%the distribution of morphological types

%Our results show a well defined framework of morphological types
%distribution in the local Universe.....

%{\it (confronto con Oesch) Oesch et al. (2010) examinating a sample of
%galaxies from COSMOS found that the evolution of morphological
%distribution in galaxies since $z\sim 1$ depends strongly on stellar
%mass with little evolution at mass above $M\sim 10^{11}$.}
\end{itemize}
\section*{Acknowledgments}
We thank Preethi Nair for sending us their data, and
the anonymous referee for a careful reading and useful comments.
BV and BMP acknowledge financial support from ASI contract I/016/07/0.
%\vspace{-1cm}

\label{lastpage}
\end{document}